\documentclass[letter]{aa} % for the letters 
\usepackage{graphicx}
\usepackage{mathtools}
\usepackage{amsmath}
\usepackage{units}
\usepackage{txfonts}
\def\lsim{\mathrel{\rlap{\lower4pt\hbox{\hskip1pt$\sim$}}
    \raise1pt\hbox{$<$}}}                % less than or approx. symbol
\def\gsim{\mathrel{\rlap{\lower4pt\hbox{\hskip1pt$\sim$}}
    \raise1pt\hbox{$>$}}}                % greater than or approx. symbol

\def\Msun{\ifmmode{\mathrm M_\odot}\else{M$_\odot$}\fi}
\newcommand{\rbreak}{\ensuremath{R_\mathrm{brk II}}}

\newcommand{\muo}{\ensuremath{\mu_\mathrm{0}}}
\newcommand{\mue}{\ensuremath{\mu_\mathrm{e}}}
\newcommand{\re}{\ensuremath{r_\mathrm{e}}}
\newcommand{\BD}{\ensuremath{B/D}}
\newcommand{\dhzdR}{\ensuremath{\nabla_{R} h_{z}}}
\newcommand{\hz}{\ensuremath{h_{z}}}
\newcommand{\hr}{\ensuremath{h_{R}}}
\newcommand{\hzhr}{$\nicefrac{h_{z}}{h_{R}}$}
\newcommand{\hi}{\ensuremath{h_{\mathrm{i}}}}
\newcommand{\ho}{\ensuremath{h_{\mathrm{o}}}}

\begin{document} 

   \title{Type-II surface brightness profiles in edge-on galaxies produced by flares}
    \titlerunning{Type II surface brightness profiles in edge-on galaxies produced by flares}
\authorrunning{Borlaff\inst{1,3}, Eliche-Moral\inst{2}, Beckman\inst{1,3,4} and Font\inst{1,3} }
%\subtitle{}

\author{Alejandro Borlaff\inst{1,3}, M.~Carmen Eliche-Moral\inst{2}, John Beckman\inst{1,3,4} and Joan Font\inst{1,3} } 

\institute{
Instituto de Astrof\'{i}sica de Canarias, C/ V\'{i}a L\'actea, E-38200 La Laguna, Tenerife, Spain, \email{asborlaff@iac.es}
\and
Departamento de Astrof\'{\i}sica y CC.~de la Atm\'osfera, Universidad Complutense de Madrid, E-28040 Madrid, Spain
\and
Facultad de F\'{i}sica, Universidad de La Laguna, Avda. Astrof\'{i}sico Fco. S\'{a}nchez s/n, 38200, La Laguna, Tenerife, Spain
\and
Consejo Superior de Investigaciones Cient\'{i}ficas, Spain
}
% \abstract{}{}{}{}{} 
% 5 {} token are mandatory
  \abstract
{Previous numerical studies had apparently ruled out the possibility that flares in galaxy discs could give rise to the apparent breaks in their luminosity profiles when observed edge-on. However the studies have not, until now, analysed this hypothesis systematically using realistic models for the disc, the flare, and the bulge.
We revisit this theme by analysing a series of models which sample a wide range of observationally based structural parameters for these three components.
Using observational data, we have considered realistic distributions of bulge-to-disc ratios, morphological parameters of bulges and discs, vertical scale heights of discs and their radial gradients defining the flare for different morphological types and stellar mass bins. The surface brightness profiles for the face-on and edge-on views of each model were simulated to find out whether the flared disc produces a Type-II break in the disc profile when observed edge-on, and if so under what conditions.
Contrary to previous claims, we find that discs with realistic flares can produce significant breaks in discs when observed edge-on. Specifically a flare with the parameters of that of the Milky Way would produce a significant break of the disc at a \rbreak\ of $\sim 8.6$ kpc if observed edge-on. Central bulges have no significant effects on the results. These simulations show that flared discs can explain the existence of many Type-II breaks observed in edge-on galaxies, in a range of galaxies with intermediate to low break strength values of $-0.25 < S < -0.1$.}

   \keywords{Galaxies: formation -- galaxies: fundamental parameters -- galaxies: evolution  --  galaxies: structure -- Galaxy: disk  --  Galaxy: structure}
   \maketitle
%
%________________________________________________________________

\section{Introduction}
\begin{figure*}[!htb]
%\vspace{-1.9cm}  %For referee version
\vspace{-2.9cm}  %For letter version 
\includegraphics[width=\textwidth]{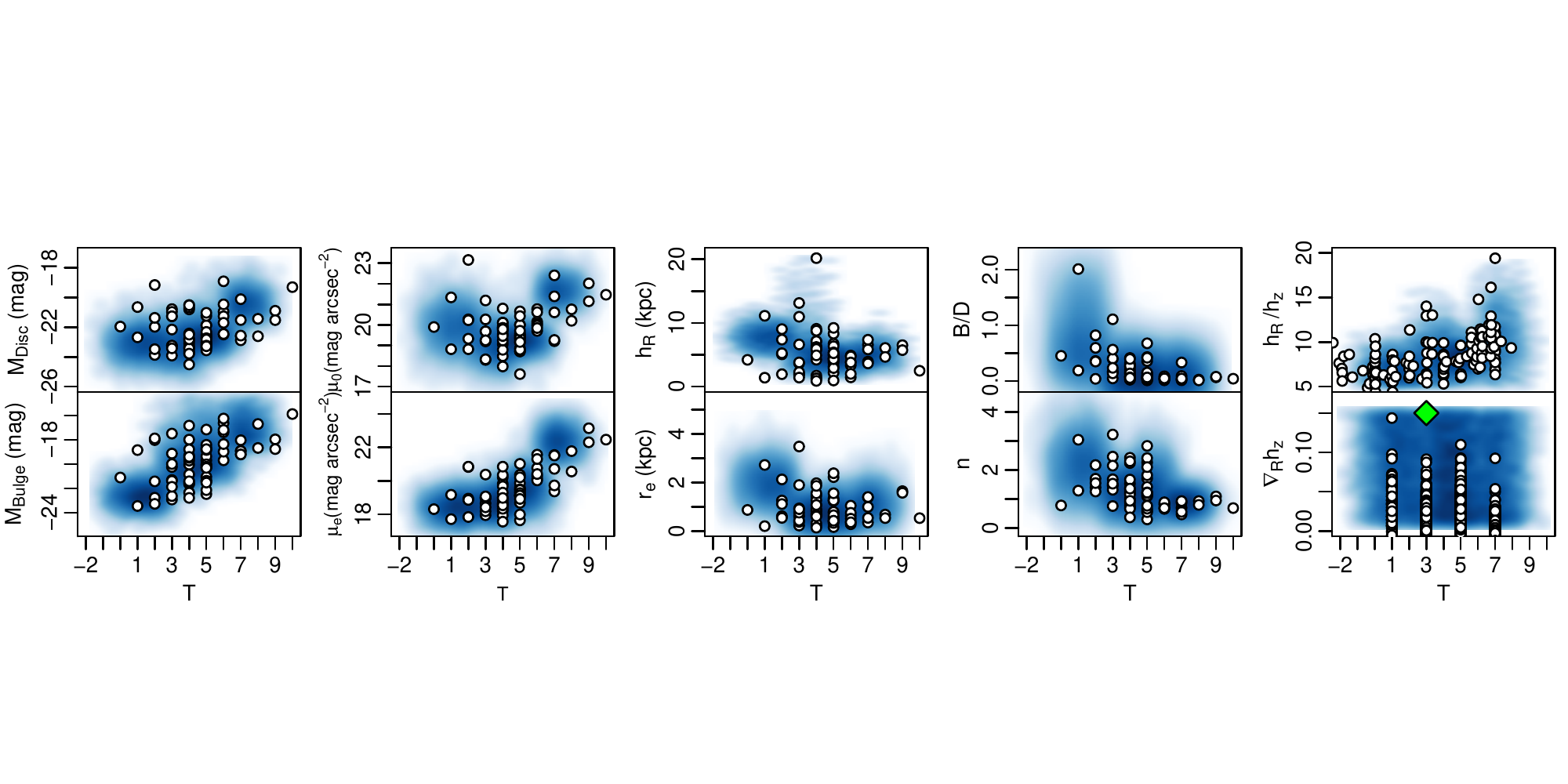}%
\vspace{-2.2cm}  %For letter version
%\vspace{-1.5cm}  %For referee version 
\caption{Photometric and structural parameter distributions for the simulations. Density clouds represent the parameter distributions of the 10,000 3D simulations of realistic bulge + flared disc light distributions created from the observations as a function of the morphological type, T. Dots represent the observations from \citet{2001AJ....121..820G,2002MNRAS.334..646KVa,2014ApJ...787...24B} and \citet{2015MNRAS.451.2376M}. \emph{Upper row, from left to right:} disc absolute magnitude $M_{abs,disc}$, disc central surface brightness \muo, disc scale length \hr, bulge-to-disc ratio \BD, and disc radial-to-vertical scale-lengths ratio \hr/\hz. \emph{Lower row, from left to right:} bulge absolute magnitude $M_{abs,bulge}$, bulge surface brightness at the bulge effective radius \mue, effective radius \re, bulge S\'{e}rsic index $n$, and the radial gradient of \hz, \dhzdR. The green diamond represents the MW.}
\vspace{-0.2cm}  %For letter version
\label{fig:Str_photo_params}
\end{figure*}

Disc galaxies often do not show a single radial exponential surface brightness profile \citep{1970ApJ...160..811F}. \citet{2006A&A...454..759P} and \citet{2008AJ....135...20E} classified the discs of galaxies into three classes according to the shape of their profiles. Type-I discs are well modelled with a single exponential profile. Type-II discs present a down-bending profile, i.e. a brightness deficit in the outer parts of the disc with respect to the extrapolation  of the inner profile outside a given ``break radius''. Finally, the profiles of Type-III discs are shallower outside the break radius than the extrapolation of the inner exponential profile (antitruncation). In the present study we focus on Type-II profiles. 

Sharp truncations in galaxy disc profiles were first reported in edge-on galaxies \citep{1979A&AS...38...15V}. Owing to projection effects, edge-on galaxies present a larger number of truncations and Type-II breaks than their face-on counterparts \citep{2007MNRAS.378..594P}. Recent studies associate sharp truncations with stellar density cut-offs and inner down-bending breaks to different stellar formation thresholds \citep{2012MNRAS.427.1102M}. \citet{2014MNRAS.441.1992L} found an interesting correlation between the break in Type-II profiles and the characteristic radii of structures in the galaxy: lenses, rings or pseudorings. This means that these morphological features may be associated with the break in some Type-II profiles. 
Previous studies also explored the possibility that flares (i.e. an increase in the scale height of the stellar disc with galactocentric radius) may be responsible for the down-bending profiles detected in highly inclined galaxies. More than a decade ago, \citet{2002MNRAS.334..646K} performed 2D decompositions of 34 edge-on late-type galaxies; 60\% of them showed Type-II profiles. The authors studied whether a linear radial increase in the scale height of the disc could produce some of the detected truncations if the discs were observed at high inclination. They created a set of artificial images ($\sim$50 simulations) using a random uniform distribution covering the observed range of values of the disc central surface brightness (\muo), disc scale length (\hr), disc scale height (\hz), bulge effective surface brightness (\mue), bulge effective radius (\re), and bulge axial ratio ($q$) of their data. They also restricted the bulge-to-disc ratio to $\BD < 2$. Finally, they concluded that values of the radial gradient of the scale height larger than 1 \hz\ per \hr\ are needed -- according to their models -- to reproduce the characteristics of the observed truncations. However, this gradient in late-type galaxies is typically very low \citep[$\dhzdR \sim 0.1\ \nicefrac{\hz}{\hr}$, see][]{1997A&A...320L..21D}, so \citeauthor{2002MNRAS.334..646KVa} concluded that flaring is a very unlikely explanation for the breaks observed in late-type galaxies.

Radial variations of the scale height of galaxy discs are difficult to measure \citep{2015MNRAS.451.2376M}, but as better and deeper data are obtained, higher values of the radial gradient of the scale height (\dhzdR) are being derived even for late-type galaxies \citep{2014ApJ...787...24B}. The Milky Way (MW, which is thought to be an SBbc galaxy) exhibits a thin+thick disc structure, both affected by a particularly strong flare \citep[$\dhzdR \sim 0.15$ at 20 kpc for the thick disc;][]{2002A&A...394..883L,2014A&A...567A.106L,2014ApJ...794...90K}. Cosmological simulations also predict much stronger flares for old stellar discs than current data reveal \citep{2015ApJ...804L...9M}. These authors have pointed out that the phenomenon may be unavoidable on the outskirts of galaxies. Consequently, the effects of flares on the disc profiles of edge-on galaxies may have systematically been underestimated in previous studies. 

No study has analysed systematically whether flares can produce Type-II profiles in highly inclined discs, using realistic models for the bulge, the disc, and (especially) the flare. So we have revisited this question to quantify the possible effects of realistic flares on the generation of breaks in edge-on disc profiles. To this end, we used observations for different morphological types and masses to create 3D models of disc galaxies assuming distributions of \BD, bulge and disc photometric parameters (\muo, \hr, \mue, \re, Sérsic index $n$), scale heights of the discs, and their trend with the galactocentric radius. To properly sample the complete range of parameters, we performed 10,000 simulations of realistic flared disc galaxies. We derived edge-on and face-on surface brightness profiles for the resulting flared galaxy models. The surface brightness profile of each galaxy model was analysed to find out whether the flare produces a significant break in the disc in the edge-on view compared with the face-on view.

\section{Methods}
We created 10,000 3D models of galaxies, each with an exponential disc plus a Sérsic bulge. We adopted the following functional form for the exponential disc \citep{2002A&A...394..883L}: 
\begin{equation} \label{eq:Corredoira}
 \rho(R,z)=\rho_{\mathrm{0}}\cdot\exp{\Bigg(\frac{-R}{h_{R}}\Bigg)}\cdot \exp{\Bigg(\frac{-|z|}{h_{z}(R)}\Bigg)}\cdot \frac{h_{z}(R)}{h_{z}(0)} \ . 
\end{equation}
Following the observations, we explored the effect of a linear increase in the vertical scale height, as follows:
\begin{equation} \label{eq:Flares_linear}
h_{z}(R) = 
\begin{cases}
h_{z}(0) & \text{if } R \leq R_{\mathrm{flare}} \\
h_{z}(0) + \dhzdR \cdot R & \text{if } R > R_{\mathrm{flare}}.
\end{cases}
\end{equation}

\citet{2001AJ....121..820G} analysed a sample of 86 face-on disc-dominated galaxies previously selected by \citet{1994A&AS..106..451D}. This author performed a bulge + disc decomposition for 69 galaxies in the $I$ band, correcting for the effects of the internal extinction, Galactic extinction, inclination, and cosmological dimming \citep{2003AJ....125.3398G}, that we used as a reference for our models. We estimated stellar masses using the relationship between the $V$-band mass-to-light ratio of galaxies and their dust-corrected rest-frame colours derived by \citet{2013MNRAS.431..430W}. According to these authors, for $z<0.1$ the optimal observed colour is $(B-V)$, so we have estimated this colour for the 69 galaxies of \citet{2003AJ....125.3398G} from HyperLeda data\footnote{HyperLeda database available at: http://leda.univ-lyon1.fr/}, and estimated the stellar mass of each galaxy using the relations in \citeauthor{2013MNRAS.431..430W} For those objects without $(B-V)$ available in HyperLeda, we estimated them from their SDSS $(g-r)$ colour following the transformations published in \citet{2005AJ....130..873J}. To simulate realistic images of the disc galaxies, we adopted the observational $I$-band distributions of \citet{2001AJ....121..820G, 2003AJ....125.3398G} for the photometric parameters (\re, S\'{e}rsic index $n$, \hr, \muo, \mue, $B/T$, the absolute magnitudes of the disc $M_{abs,disc}$ and the bulge $M_{abs,bulge}$), and four morphological type bins (S0--Sa, Sb--Sbc, Sc--Scd and Sd--Sdm), in three mass bins ($10 < \log_{10}M/M_{\odot} < 10.7, 10.7 < \log_{10}M/M_{\odot} < 11$ and $\log_{10}M/M_{\odot} > 11$) in order to explore realistic mass distributions for the morphological type bins. In Fig.\ref{fig:Str_photo_params} we represent the distributions of the structural and photometric parameters from which we created the models and compared them to the observations they are based on. For each morphological type bin we randomly chose the ratio of scale height to scale length (\hzhr) from the observational range of values corresponding to each type reported by \citet{2002MNRAS.334..646K} and \citet{2015MNRAS.451.2376M} in the $I$ band, as shown in Fig.\,\ref{fig:Str_photo_params}.

\begin{figure}
\includegraphics[angle=0,width=0.95\linewidth]{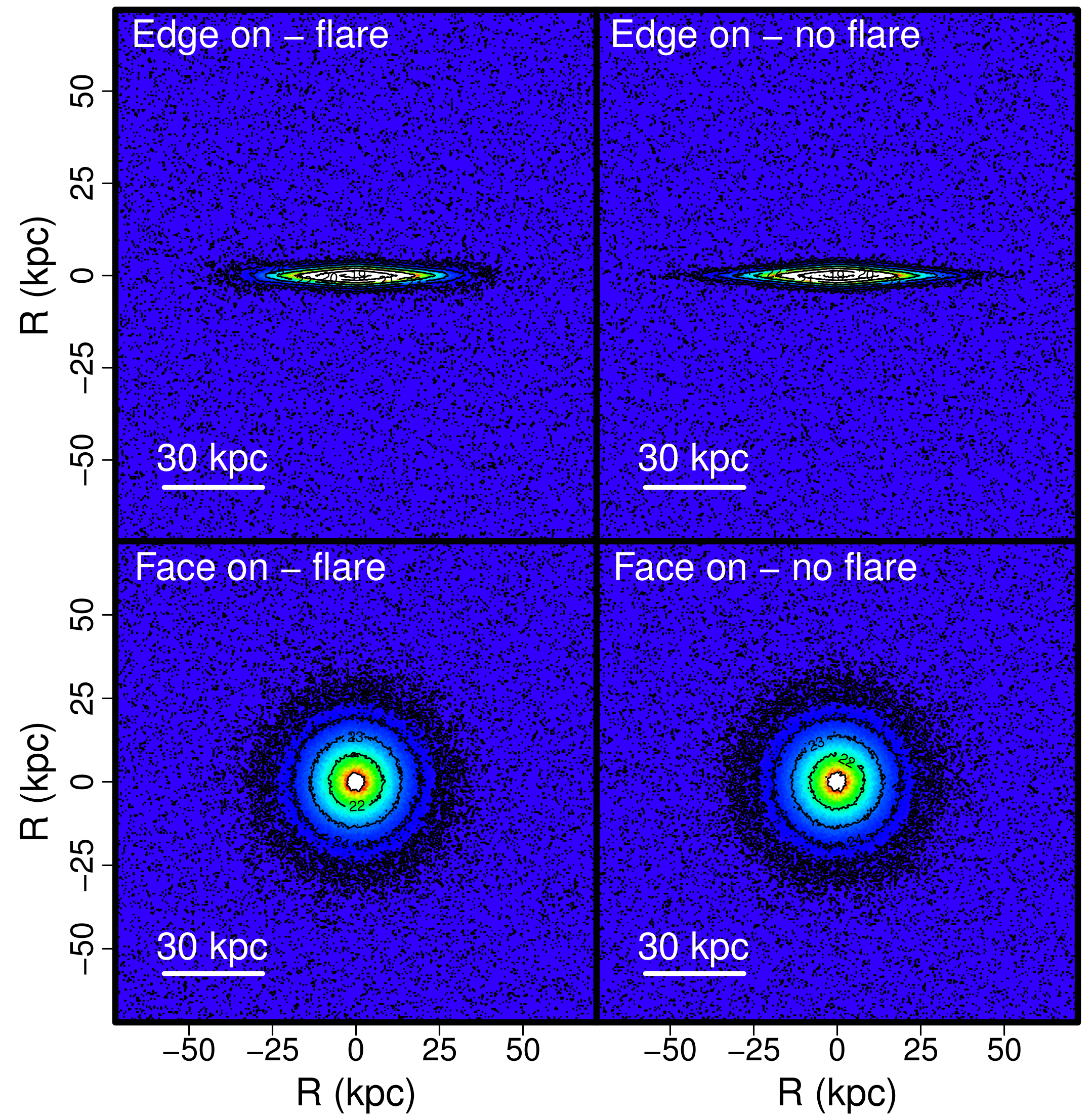}\\
\includegraphics[angle=0,width=0.95\linewidth]{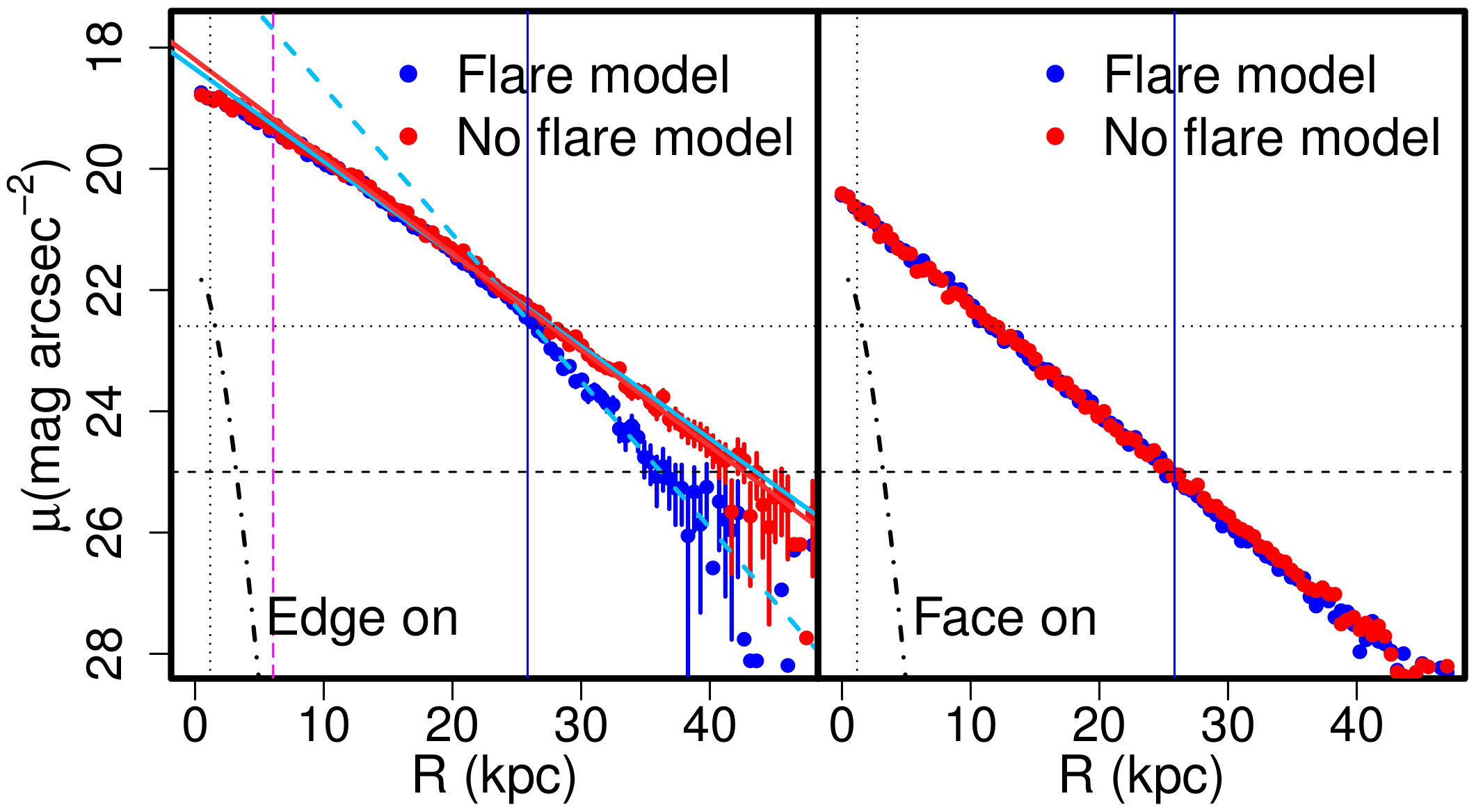}
\caption{\textbf{Top panels:} Edge-on $I$-band images for the flared and non-flared models. \textbf{Intermediate panels:} The same for the face-on discs. \textbf{Bottom panels:} Surface brightness profiles in the major axis of the galaxy for edge-on (\emph{left}) and the face-on (\emph{right}) views of the two models (\emph{red}: non-flared model, \emph{blue}: flared model). \emph{Blue vertical solid line}: flare radius $R_{\mathrm{flare}}$. \emph{Black dot-dashed line}: bulge component. \emph{Solid blue and red lines}: linear fits performed to the inner regions of the disc for the flared and non-flared models. \emph{Dashed blue and red lines}: The same for the outer regions of the disc. \emph{Black dotted lines}: Bulge \re\ and \mue values. \emph{Horizontal dashed black line}: Limiting magnitude $\mu_{lim}=25$ mag arcsec$^{\mathrm{-2}}$. \emph{Vertical dashed magenta line}: Inner limit for the inner profile. We note that the disc profile in the flared case exhibits a clear Type-II break in the edge-on view, whereas the non-flared model keeps the exponential profile that both models exhibit in the face-on view. See the legend for details.}
\label{fig:Ima_profiles}
\vspace{-0.5cm}
\end{figure}

We explored the range of \dhzdR\ values derived by recent observational studies for galaxies of different types (see also Fig.\,\ref{fig:Str_photo_params}) in the $I$ band: $0<\dhzdR<0.15$ \citep{2014ApJ...787...24B,2015MNRAS.451.2376M}. These values agree with those measured in the MW by several authors \citep[and references therein]{2000astro.ph..7013A,2002A&A...394..883L,2006A&A...451..515M,2011A&A...527A...6H,2014A&A...567A.106L,2014ApJ...794...90K}. \citet{2015MNRAS.451.2376M} argued that the scale heights derived from 1D decompositions may be biased to higher values than reality owing to bulge contamination (up to 10\% in galaxies with $B/T\sim0.4$). Therefore, we restricted the sample from which we obtain the scale-height gradients to objects with $B/T < 0.4$ to avoid bulge contamination. The radius at which the flare starts ($R_{\mathrm{flare}}$) is randomly chosen following a uniform distribution between 2 and 5 times the scale length of the disc. The lowest value is set to avoid bulge contamination from the inner regions of the profile \citep{1997A&A...320L..21D}, while the highest value is the typical maximum radius where the limiting magnitude of both edge-on and face-on images is reached: $\mu_{lim}=25$ mag arcsec$^{\mathrm{-2}}$. At higher radial distances the flare would not have any significant effect on our profiles. Finally, once a set of parameters describing a realistic galaxy model is created, we check that the chosen values of  $M_{abs,disc}$ and $M_{abs,bulge}$ provide a $B/T$ ratio within the observed distribution of $B/T$ values for the stellar mass and morphological type of the galaxy. We created a non-flared analogue of each flared model in order to compare their edge-on profiles.
The 3D models of luminosity distribution in the $I$ band have a spatial resolution of 1 arcsec, with a total size of 301 arcsec in each direction. The adopted resolution is similar to the typical FWHM achieved with ground-based seeing-limited observations. We assumed a distance of $100$ Mpc to the simulated object, which is the upper limit of the distance distribution of the galaxies in \citet{2001AJ....121..820G}. Owing to the increase in the thickness of the disc in the flared models, the simulations might suffer from flux loss because of the limited box size. To prevent this, we automatically removed any models with differences greater than 1\% in total flux between the flared and non-flared models. We created face-on and edge-on density images by integrating along the line of sight. The images are then convolved with a Gaussian kernel of $\sigma = 1$ arcsec. The surface brightness profile for each galaxy model has been derived for edge-on and face-on views, by locating a  1 arcsec wide single slit along the major axis in the first case and with the {\tt{ellipse}} task in IRAF in the second. 
We included Poissonian noise across the whole profile. We assumed Planck cosmology when required \citep[$\Omega_{M} = 0.31, \Omega_{\Lambda}=0.69$, H$_{\ensuremath{0}}$ = 67.8 km s$^{\ensuremath{-1}}$ Mpc$^{\ensuremath{-1}}$;][]{2015arXiv150201589P}.

We took the model of the MW presented in \citet{2014A&A...567A.106L} as reference. These authors fitted SDSS-SEGUE MW data to a flared thin+thick disc model, obtaining compatible and deeper results than previous studies \citep{2002A&A...394..883L}. They provide the 3D distribution, \hz\ distribution, as well as the fitted parameters for the thin and thick discs and both flares. We built up this model following the described procedure. For this model, we also created a non-flared analogue MW model by assuming the scale height at the sun's radius to be constant for the whole disc.

\section{Results and discussion}

In Fig.\,\ref{fig:Ima_profiles} we present one example of the realistic flared and non-flared models that we have created. The top and middle panels represent the simulated images for the flared and non-flared model of a galaxy with a stellar mass of $10^{\ensuremath{10}} M_{\odot}$ and type Sd-Sdm, for edge-on and face-on views. The lower panels show the surface brightness profiles for the two models and for both inclinations (see the caption for details). The main result is that there is no significant difference between the flared and non-flared models in face-on orientation, either in the images or in the surface brightness profiles. On the contrary, when we compare the models in edge-on orientation, we clearly detect a break in the surface brightness profile of the flared model that does not appear in the non-flared analogue. 
We have found detectable breaks in the edge-on profiles of $\sim42\%$ of all our models including a flare, with a mean strength\footnote{The strength of a disc break is defined as $S=\log_{10}{(\ho/\hi)}$, with \hi\ and \ho\ the scale lengths of the inner and outer discs.} of $S=-0.15$, and 98\% of them with $-0.21 < S < -0.11$. We obtain higher detection rates in brighter discs (i.e. higher masses), and in those models with values for $\dhzdR$ greater than $\sim 0.05$. This is expected if a fixed limiting magnitude and spatial resolution is assumed. This means that realistic flares can frequently produce noticeable Type-II breaks in highly inclined galaxies, contrary to previous claims.

Concerning the MW reference model, we find a significant drop in its surface brightness profile at $8.58^{+0.43}_{-0.31}$ kpc in the edge-on profile in the flared model, which is not observed either in the face-on profile or in the edge-on profile of its non-flared analogue. We obtained a break strength $S=-0.22$ for the break of the MW flared model. Therefore, we conclude that a flare such as the one reported for the MW in many studies, would produce a Type-II profile if the MW were observed edge-on.

%\begin{figure*}
\begin{figure}
\vspace{-0.5cm}
\centering
\includegraphics[angle=0,width=0.94\linewidth]{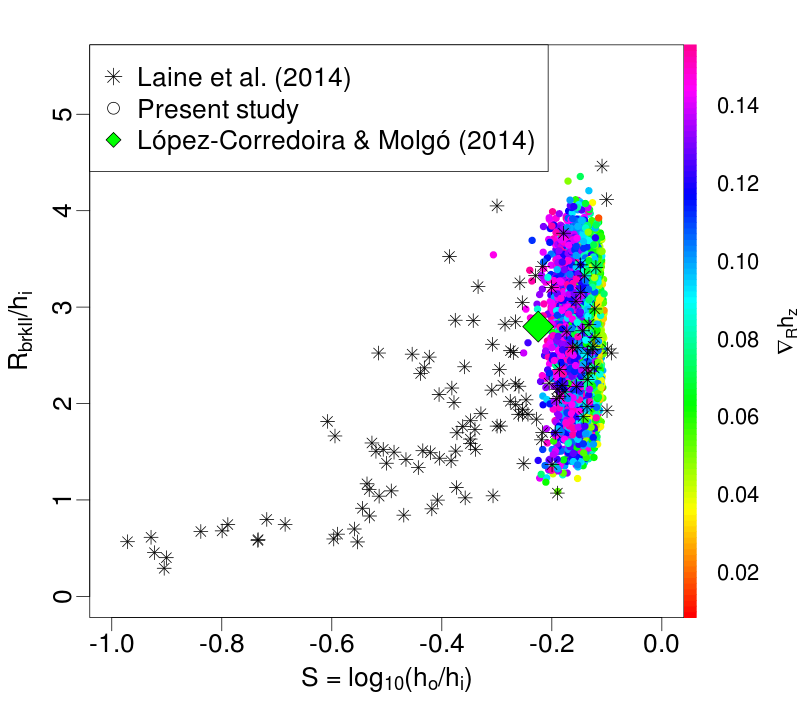}
\caption{Break radius normalized by the inner disc scale length vs. the break strength $S$. Our simulations are plotted with colour-coded circles according to \dhzdR. The edge-on MW model is overplotted with a green diamond. The asterisks represent the Type-II breaks observed by \citet{2014MNRAS.441.1992L} from the S$^{4}$G survey. See the legend in the figure.}
\vspace{-0.3cm}
\label{fig:Laine_comp}
%\end{figure*}
\end{figure}

\begin{figure}[!t]
\includegraphics[angle=0,width=0.94\linewidth]{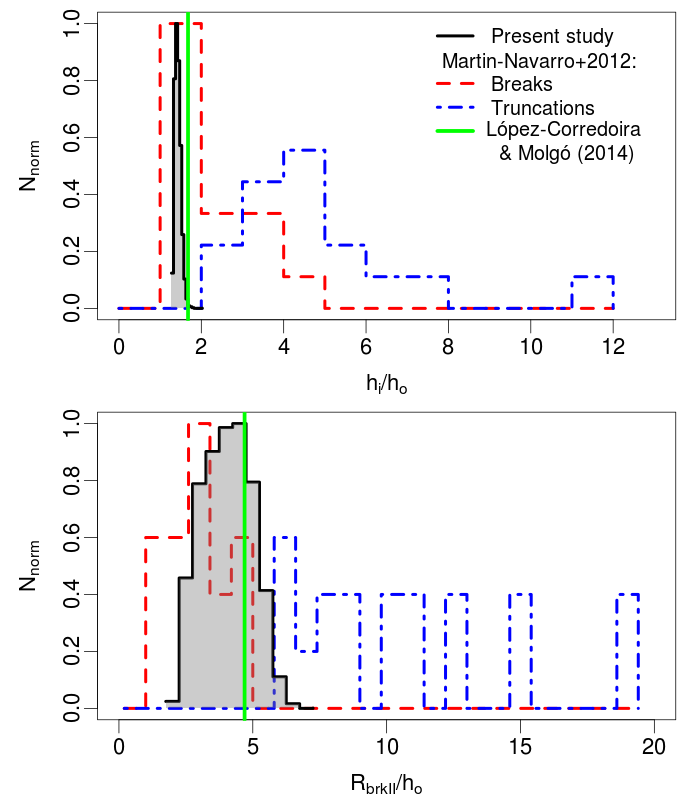}
\caption{Histograms comparing the distribution of Type-II profiles from the present study and \citet{2012MNRAS.427.1102M}, as a function of the ratio between the inner and the outer scale length  \hi/\ho\ (upper panel) and of the break radius normalized by the outer scale length \rbreak/\ho\ (lower panel). The vertical green lines mark the values for the MW reference model. See the legend in the figure.}
\vspace{-0.2cm}
\label{fig:MN2012_comp}
\end{figure}

In Fig.\,\ref{fig:Laine_comp} we represent the ratio between the break radius and the scale length of the inner disc ($\rbreak / \hi$) vs. the break strength ($S$). The flared models reproduce the observations by \citet{2014MNRAS.441.1992L} in the lower break strength part of the real distribution ($-0.25<S<-0.1$). We note that our 3D models are dust-free by construction. We can compare them because the authors use data in the $K$ and 3.6 microns bands; these data are negligibly affected by dust extinction. We find a significant correlation between $S$ and \dhzdR\ ($\rho_{Spearman} = -0.32,$ p-value $< 10^{\mathrm{-5}}$), which is not found between $S$ and the morphological type (see Fig. A.1, available in the online version). 

In Fig.\,\ref{fig:MN2012_comp}, we compare characteristics of the Type-II profiles from the observational sample by \citet{2012MNRAS.427.1102M} with those of our flared models. These authors identify two types of Type-II profiles: the \emph{breaks}, which appear as down-bending transitions, and the \emph{truncations}, much steeper and located much further out. In order to reproduce the latter type, often classified as \emph{sharp} truncations, we would require significantly greater values of \dhzdR\ than those reported in galaxies by current observational studies, including the MW. This contrasts with the \emph{breaks}, whose distribution can be partially reproduced by flared discs, as demonstrated by our models. 

We have also studied the surface brightness profiles as a function of the distance from the galactic plane in the flared models that exhibit breaks in their edge-on profiles. We find an increase in the values of \hi\ and \ho\ with $z$ equal to factor of $1.15^{+0.23}_{-0.12}$ for the inner disc and $2.19^{+1.21}_{-0.72}$ for the outer disc up to a maximum height of $z/\hz \leq 5$, such that \ho\ becomes closer to \hi. This agrees with the results of \citet{2007MNRAS.378..594P}, where the authors found a significant weakening of the breaks with increasing distance from the plane for both inner and outer profiles, with factors of 1.1-1.4 and 1.4-2.4, respectively. \emph{So we can conclude that the Type-II discs produced by flares in our simulations can reproduce the global properties of real edge-on Type-II discs with low to intermediate break strengths.}
\section{Conclusions}
The present realistic 3D models of flared disc galaxies demonstrate that, contrary to previous claims, flares can produce Type-II profiles in galaxies when viewed edge-on, especially for breaks with intermediate to low strength ($-0.25<S<-0.05$). We also find a significant weakening of the breaks with the distance from the plane, which agrees with observations. We do not find any correlation between the existence of a significant bulge component and the presence of a Type-II break. 

\begin{acknowledgements}
The authors thank to the anonymous referee for the useful criticisms that helped to improve this publication significantly and Martín López-Corredoira for his kind support with the MW models. This research has been supported by the Ministerio de Economía y Competitividad del Gobierno de España (MINECO) under project AYA2012-31277, and by the Instituto de Astrofísica de Canarias under project P3/86. We acknowledge the usage of the HyperLeda database (http://leda.univ-lyon1.fr). This research has made use of the NASA Astrophysics Data System and NASA/IPAC Extragalactic Database (NED) and the $R$ environment for statistical computing.  
\end{acknowledgements}
%-------------------------------------------------------------------
\bibliographystyle{aa}
\bibliography{borlaff_14.bib}{}

\newpage
\section{Online version material}
\setcounter{figure}{0} 
\renewcommand{\thefigure}{A.\arabic{figure}} 
%\counterwithin{figure}{Additional content}
\begin{figure}[!h]
\vspace{-0.5cm}
\centering
\includegraphics[angle=0,width=0.99\linewidth]{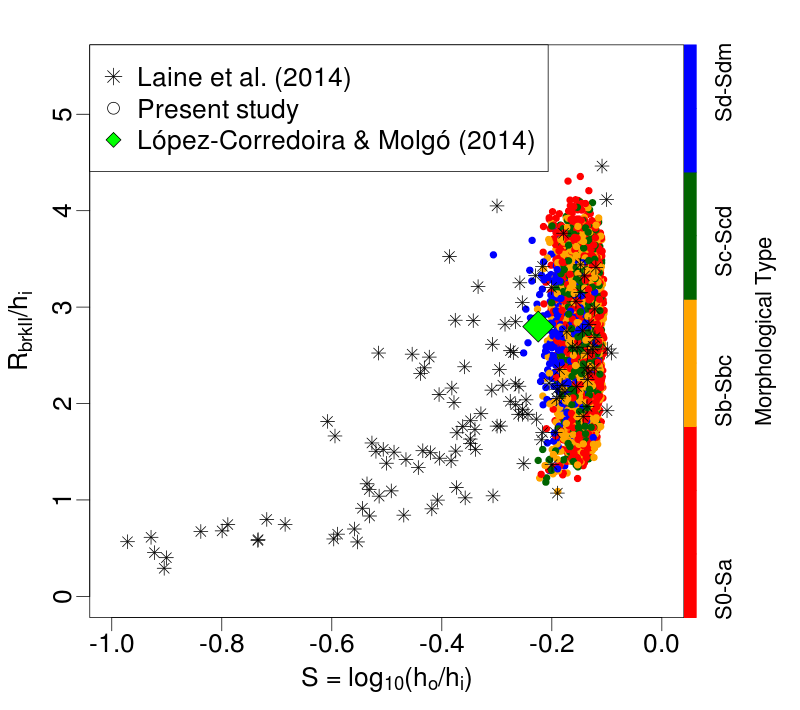}
\caption{Break radius normalized by the inner disc scale length vs. the break strength $S$. Our simulations are plotted with coloured circles, colour-coded according to the morphological type of the object simulated object. The edge-on MW model is overplotted with a green diamond. The asterisks represent the Type-II breaks observed by \citet{2014MNRAS.441.1992L} from the S$^{4}$G survey. See the legend in the figure. [Figure available in the online version.]}
\vspace{-0.3cm}
\label{fig:Laine_comp_2}
\end{figure}

\end{document}